\documentclass{article}

\usepackage{graphicx}
\usepackage{amssymb}
\usepackage{amsmath}
\usepackage{color}
\usepackage{url}
\usepackage{graphicx}        

\newenvironment{sbmatrix}[1]{\left[ \begin{array}{#1}}%
	{\end{array} \right]}

\newcommand{\VYL}{\ensuremath{V^Y_\Lambda}}

\newcommand{\KL}{\ensuremath{K^L}}
\newcommand{\VL}{\ensuremath{V^L}}
\newcommand{\KY}{\ensuremath{K^Y}}
\newcommand{\VY}{\ensuremath{V^Y}}
\newcommand{\KYL}{\ensuremath{K^Y_\Lambda}}
\newcommand{\KDR}{\ensuremath{K^{DR}}}
\newcommand{\VDR}{\ensuremath{V^{DR}}}

\newcommand{\bfr}{\ensuremath{\mathbf{r}}}
\newcommand{\bfrprime}{\ensuremath{\mathbf{r}^\prime}}

\newcommand{\normderiv}[1]{\ensuremath{\frac{\partial #1}{\partial n}}}

\def\XXint#1#2#3{{\setbox0=\hbox{$#1{#2#3}{\int}$}
     \vcenter{\hbox{$#2#3$}}\kern-.5\wd0}}

\usepackage{fullpage}

\begin{document}

\title{Analytical Nonlocal Electrostatics Using Eigenfunction Expansions of Boundary-Integral Operators}
\author{Jaydeep P. Bardhan\\
Dept. of Molecular Biophysics and Physiology, Rush University Medical Center\\
\and
Matthew G. Knepley\\
Computation Institute, University of Chicago\\
\and
Peter Brune\\
Mathematics and Computer Science Division, Argonne National Laboratory}

\maketitle

\begin{abstract}
In this paper, we present an analytical solution to nonlocal continuum
electrostatics for an arbitrary charge distribution in a spherical
solute.  Our approach relies on two key steps: (1) re-formulating the
PDE problem using boundary-integral equations, and (2) diagonalizing
the boundary-integral operators using the fact their eigenfunctions
are the surface spherical harmonics.  To introduce this uncommon
approach for analytical calculations in separable geometries, we
rederive Kirkwood's classic results for a protein surrounded
concentrically by a pure-water ion-exclusion layer and then a dilute
electrolyte (modeled with the linearized Poisson--Boltzmann equation).
Our main result, however, is an analytical method for calculating the
reaction potential in a protein embedded in a nonlocal-dielectric
solvent, the Lorentz model studied by Dogonadze and Kornyshev.  The
analytical method enables biophysicists to study the new nonlocal
theory in a simple, computationally fast way; an open-source MATLAB
implementation is included as supplemental information.
\end{abstract}

\section{Introduction}\label{sec:introduction}
One of the long-standing challenges in molecular biophysics is the
development of accurate, yet simple models for the influence of
biological fluids (aqueous solutions composed of water and dissolved
ions) on biological molecules such as proteins and DNA.  Atomistic
simulations that include explicit water molecules, such as molecular
dynamics (MD), provide the most detailed molecular understanding that
is widely accessible without specialized computational resources.
However, these simulations come at two prices: first, MD simulations
can require many hundreds of compute hours, most of which are spent on
the thousands of water molecules whose individual behaviors are not of
primary relevance; second, practitioners must understand numerous
subtleties about simulation protocols and the parameters associated
with the physical models (force fields).  Implicit-solvent models
replace the explicit water molecules with an approximation to the
theoretically rigorous potential of mean force (PMF)~\cite{Roux99},
creating the possibility of simulating molecular behavior accurately
but orders of magnitude faster, and with fewer statistical
uncertainties.  Unfortunately, the statistical mechanical derivation
of the PMF is not constructive, in the sense that the derivation does
not provide a general PMF suitable for all molecular solutes.
Instead, one must guess a functional form, such as the Poisson
equation for the electrostatic interactions between solvent and
solute, find the optimal parameters, and then test its fit against
real data (both experiment and more accurate theories such as MD).

Of course, evaluation of an implicit-solvent model is greatly
accelerated if it can be solved easily and rapidly on relevant,
non-trivial problems.  With the advent of fast computers, one
reasonable option is to make numerical software implementing the new
model freely available online~\cite{Azuara08,Koehl10}.  Another option
is to provide analytical solutions for tractable geometries.  Spheres
are frequently used for continuum electrostatic modeling, because
exact results can be obtained using spherical harmonics and the method
of separation of variables~\cite{Kirkwood34,Jackson_classical_electrodynamics}.  Kirkwood's
classic solution for a spherical protein embedded in a dilute
electrolyte represents the best-known example~\cite{Kirkwood34}, and
demonstrates this conceptually simple approach.  One merely writes
down spherical-harmonic expansions and matches expansion coefficients
using the known boundary conditions.  Even though proteins obviously
have complicated shapes, analysis of spherical geometries can offer
insights into problems such as p$K_\mathrm{a}$
predictions~\cite{Havranek99}, redox potentials~\cite{Zhou97_2},
strategies for optimizing molecular binding~\cite{Kangas98}, and fast
analytical models such as Generalized Born~\cite{Sigalov05}.

However, Kirkwood's work also demonstrates a difficulty with the
approach: as one adds detail to the model---in Kirkwood's case, an
ion-exclusion layer outside the protein---calculations become
onerously complex very quickly.  Every additional layer or unknown
function introduces another set of expansions that need to be matched,
and manual algebraic manipulation for the desired expansion
coefficients essentially entails solving a linear system of equations,
so that the number of operations grows cubically with the number of
equations.  In addition, modeling the linearized Poisson--Boltzmann
equation in the solvent necessitated the introduction of a set of
polynomials for the radial coordinate because the standard Bessel
functions were unsuitable~\cite{Kirkwood34}; more than sixty years
passed before the relationship between Kirkwood's polynomials and the
Bessel functions was established, allowing at the end a substantial
simplification~\cite{Mladenov96}.

In this paper, we present an alternative strategy for obtaining
analytical solutions in separable geometries.  The first step is to
transform the given system of partial-differential equations (PDEs)
into one of boundary-integral equations (BIEs)~\cite{Atkinson97}, so
that the unknowns are no longer functions defined over
three-dimensional regions of space, but instead functions defined on
two-dimensional boundaries.  Second, the boundary-integral operators
are diagonalized using the appropriate
harmonics~\cite{Bowman,Ritter95_spectrum}.  This allows a mode-by-mode
calculation of the unknown functions on the boundary in terms of the
appropriate \textit{surface} harmonics---in contrast to
matched-expansion approaches that employ \textit{solid} harmonics.  To
demonstrate the BIE-eigenfunction approach, we solve the Kirkwood
problem (a spherical protein embedded in a dilute electrolyte, with a
thin ion-exclusion or Stern layer~\cite{Kirkwood34}) and derive the
full solution to the more recent nonlocal-dielectric model of
Dogonadze and Kornyshev~\cite{Dogonadze74,Kornyshev78,Vorotyntsev78}.

The nonlocal model was originally developed to address one of the key
shortcomings of macroscopic continuum theories for molecular
solvation: the fact that the solvent molecules (usually water) are not
infinitesimally small compared to length scales of interest, e.g.,
small ions~\cite{Basilevsky96,Hildebrandt04} and
proteins~\cite{Rubinstein04}.  Unfortunately, nonlocal response means
that even the simplest form of the nonlocal model, called the Lorentz
nonlocal theory~\cite{Attard90}, leads to an integrodifferential
Poisson equation, which is difficult to solve analytically or even
numerically.  Consequently, to date the only analytically solved
geometries for the Lorentz nonlocal model have been the sphere with
central charge~\cite{Basilevsky98,Weggler_thesis} and the charge near
a half-space~\cite{Rubinstein04,Rubinstein07,Rubinstein10}, and no
numerical algorithms for the original nonlocal model in arbitrary
geometries were ever presented.

Very recently, however, Hildebrandt and collaborators derived several
mathematical reformulations to render the Lorentz nonlocal
electrostatic model tractable both analytically and
computationally~\cite{Hildebrandt04,Hildebrandt05,Hildebrandt07,Weggler10}.
The first major step was reformulating the nonlocal
integrodifferential Poisson problem in one unknown variable, the
electrostatic potential $\varphi(\mathbf{r})$, as a pair of coupled,
purely local PDEs with two unknown variables throughout space
($\varphi(\mathbf{r})$ and an additional auxiliary
potential)~\cite{Hildebrandt04}.  Similar reformulations of nonlocal
continuum theory were obtained independently in other areas of
physics~\cite{Ochs98,Engelen03}.  Following this reformulation,
Green's theorem and double reciprocity can be used to transform the
coupled PDE system into a purely boundary-integral-equation (BIE)
representation of the nonlocal model~\cite{Hildebrandt07,Fasel08}.

In principle, both the local-formulation PDE problem and the purely
BIE method are solved problems numerically, in the sense that
asymptotically optimal (linear-scaling) numerical algorithms
exist~\cite{Baker01,Boschitsch02,Lu06,Altman09,Weggler10,Bardhan11_DAC}.
However, even ``fast solvers'' can require an hour or more of
computation, and therefore analytical solutions of non-trivial
problems still hold significant value in this relatively early stage of
testing nonlocal electrostatics of molecular solvation.  One
application of analytical methods is to obtain qualitative insight
into the differences between nonlocal and local models using
visualization: analytical methods allow rapid calculations of the
reaction potential induced throughout a model geometry by a chemical
group in the protein, e.g. an amino acid side chain.  Another
application of analytical methods is to obtain quantitative
information that may help to determine model parameters.  For
example, the nonlocal model includes an additional parameter beyond
those of the standard local model.  This parameter, denoted by
$\lambda$, is an effective length scale that captures water's
transition from behaving like a low-dielectric material at short
length scales to more familiar high-dielectric, bulk-like behavior at
longer length scales.  Parameterization requires extensive simulation
and testing, and fast calculations aid significantly.

To support the development and testing of nonlocal electrostatic
models for biomolecule solvation, we present here the nonlocal-model
analogue of Kirkwood's result: namely, an analytical approach for the
electrostatic solvation free energy of an arbitrary charge
distribution in a spherical solute embedded in a solvent modeled as a
Lorentz nonlocal dielectric.  Kirkwood's classic work continues to
have impact decades after the advent of numerical simulations of the
continuum electrostatic model~\cite{Warwicker82,Havranek99,Sigalov05},
and the present work significantly enlarges the scope of nonlocal
problems that can be studied analytically.  We note that mobile ions
such as sodium and potassium play crucial physiological roles and that
the present work addresses only pure water solvent.  However, the
nonlocal theory can be extended easily to linearized
Poisson--Boltzmann treatment of physiological electrolyte
solutions~\cite{Hildebrandt04}, and these extensions are the subject
of ongoing work.

The remainder of the paper is organized as follows: the next section
describes the local and nonlocal models, their reformulation as
systems of boundary-integral equations, and the eigendecompositions of
the associated boundary-integral operators.  In
Section~\ref{sec:bie-eigenfunctions} we introduce our
BIE-eigenfunction strategy by rederiving the solution to Kirkwood's
problem, and then apply the strategy to solve the nonlocal problem.
In Section~\ref{sec:results}, we present several applications of the
analytical solution, which illuminate important differences between
local and nonlocal electrostatics, including the choice of solute
dielectric constant and the sensitivity of the nonlocal results to the
solvent length-scale parameter $\lambda$.  The paper concludes in
Section~\ref{sec:discussion} with a brief summary and discussion.

\section{Background}
\begin{figure}[ht!]
  \centering \resizebox{6.0in}{!}{\includegraphics{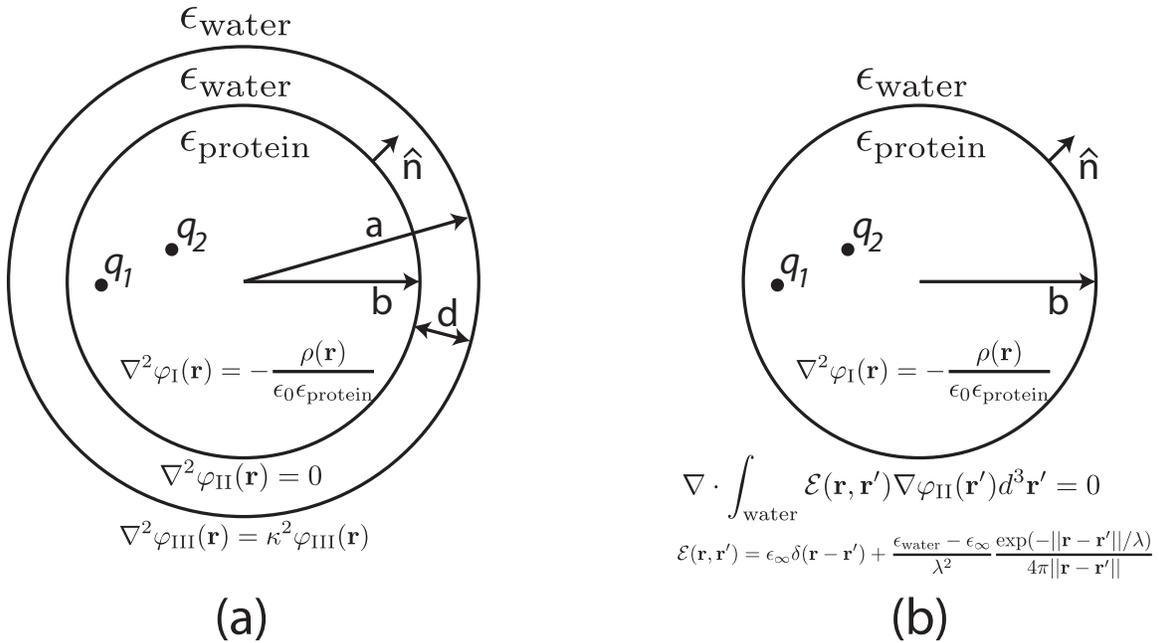}}

  \caption{Diagram of the two continuum electrostatic models to be
    solved analytically. (a) Kirkwood's problem~\cite{Kirkwood34}. (b)
    Nonlocal-response model in a pure-water
    solvent.}\protect\label{fig:problems} \end{figure}

\subsection{Kirkwood's Local-Response Electrostatic Model}
Figure~\ref{fig:problems}(a) is an illustration of the local-response
model under consideration.  We assume that the solute region~I is a
sphere of radius $b$, which is centered at the origin, and that the
solute is at infinite dilution in a dilute aqueous electrolyte
solvent.  The solute charge distribution $\rho(\mathbf{r})$ is modeled
as a set of $Q$ discrete point charges contained within the sphere,
the $i$th of which has value $q_i$ and is situated at $(r_i, \theta_i,
\phi_i)$.  The solute is treated as a homogeneous local-response
dielectric with relative permittivity $\epsilon_{\mathrm{protein}}$,
i.e. inside the protein, the constitutive relation between the
displacement and electric field is
\begin{equation}
  \mathbf{D}_{\mathrm{I}}(\mathbf{r}) = \epsilon_{\mathrm{protein}}\epsilon_0 \mathbf{E}_\mathrm{I}(\mathbf{r})
  \end{equation}
where as usual $\mathbf{E}(\mathbf{r}) = -\nabla \varphi(\mathbf{r})$
with $\varphi$ the electrostatic potential.  Substituting this
constitutive relation into Gauss's law for dielectrics
\begin{equation}
  \nabla \cdot \mathbf{D}_\mathrm{I}(\mathbf{r}) =
\rho(\mathbf{r}),
\end{equation}
we see the electrostatic potential in region~I satisfies the familiar Poisson equation
\begin{equation}
  \nabla^2 \varphi_\mathrm{I}(\mathbf{r}) = -\frac{\rho(\mathbf{r})}{\epsilon_0\epsilon_{\mathrm{protein}}}.
  \end{equation}
In a thin solvent layer surrounding the protein, we have water but no
mobile ions; assuming that they are point charges in hard spheres of
radius $d$, the ion density must be zero for $||\mathbf{r}|| < b + d$.
Consequently, in this region (labeled II in Figure~\ref{fig:problems}(a)) the
potential satisfies a Laplace equation and we assume the permittivity
is just that of pure water $\epsilon_\mathrm{water} \approx 80$.
Standard boundary conditions hold at the protein--solvent interface
defined by $||\mathbf{r}||=b$, namely the continuity of the potential
and the normal component of the displacement field:
\begin{align}
  \varphi_\mathrm{I}(\mathbf{r}_b^-) & = \varphi_\mathrm{II}(\mathbf{r}_b^+) \label{eq:potential-bc}\\
  \mathbf{\hat{n}} \cdot \mathbf{D}_\mathrm{I}(\mathbf{r}_b^-) & =
  \mathbf{\hat{n}} \cdot \mathbf{D}_\mathrm{II}(\mathbf{r}_b^+).\label{eq:displacement-bc}
\end{align}
For local-response dielectrics, Eq.~\ref{eq:displacement-bc} reduces to the familiar
\begin{equation}
  \epsilon_{\mathrm{protein}} \frac{\varphi_\mathrm{I}(\mathbf{r}_b^-)}{\partial n} =
  \epsilon_{\mathrm{water}} \frac{\varphi_\mathrm{II}(\mathbf{r}_b^+)}{\partial n}.
  \end{equation}
where the superscripts $-$ and $+$ denote the interior (solute) and
exterior (solvent) regions, respectively, and the normal direction
$\hat{\mathbf{n}}$ points outward from region~I to region~II.

Outside this ion-exclusion layer, the mobile ions are assumed to
redistribute such that at any point $\mathbf{r}$, the net charge
density is the sum of the Boltzmann-weighted ion densities (i.e.,
neglecting the ion sizes and correlations between them).  This leads
to the nonlinear Poisson--Boltzmann equation, which here we simplify
by linearization, i.e. the potential in region III satisfies the
linearized Poisson--Boltzmann equation (LPBE)
\begin{equation}
  \nabla^2 \varphi_\mathrm{III}(\mathbf{r}) = \kappa^2 \varphi_\mathrm{III}(\mathbf{r})\label{eq:LPBE}
  \end{equation}
where $\kappa$ is the inverse Debye screening length; for
physiological solutions, $\kappa \approx 8$~\AA.  The electrolyte is
also assumed to have relative permittivity $\epsilon_\mathrm{water}$,
and so the boundary conditions at the ion-exclusion boundary
$||\mathbf{r}||=a$ are
\begin{align}
  \varphi_\mathrm{II}(\mathbf{r}_a^-) & = \varphi_\mathrm{III}(\mathbf{r}_a^+) \\
  \frac{\varphi_\mathrm{II}(\mathbf{r}_a^-)}{\partial n} &=
  \frac{\varphi_\mathrm{III}(\mathbf{r}_a^+)}{\partial n} 
  \end{align}

Kirkwood solved the above problem for the potential using matched
expansions in the solid spherical harmonics~\cite{Kirkwood34}.  Here,
we show that an alternative is to use the surface harmonics for the
BIE formulation of this problem, which may be derived as follows.  For
a point $\mathbf{r}$ in one of these regions, Green's representation
theorem allows the potential at $\mathbf{r}$ to be written in terms of
the potential and its normal derivative at the surface or surfaces the
bound the region~\cite{Jackson_classical_electrodynamics,Juffer91,Yoon90,Altman09}.  In
region~I, for example,
\begin{align}
  \varphi_\mathrm{I}(\mathbf{r})  = \int_b \frac{\partial G^L(\mathbf{r},\mathbf{r}')}{\partial n} \varphi_\mathrm{I}(\mathbf{r}') dA'
        - \int_b G^L(\mathbf{r},\mathbf{r}') \frac{\partial \varphi_\mathrm{I}(\mathbf{r}')}{\partial n} dA'
        + \int_{\mathrm{region~I}} G^L(\mathbf{r},\mathbf{r}') \rho(\mathbf{r}') dV',\label{eq:Green-basic}
  \end{align}
where the subscript $b$ denotes the spherical boundary $||\mathbf{r}||
= b$, $G^L(\mathbf{r},\mathbf{r}') = \frac{1}{4 \pi
  ||\mathbf{r}-\mathbf{r}'||}$ is the free-space Green's function for
the Laplace equation, and the third term on the right-hand side
represents the Coulomb potential induced by the solute charge
distribution.  Writing similar expressions for the potential in
regions~II and~III, and taking careful limits as the field points
approach these bounding surfaces, we obtain a system of four
boundary-integral equations for the four unknown functions (the
potential and normal derivative on the two boundaries).  The complete
derivation is presented elsewhere~\cite{Altman09}, but the final system
may be written as
\begin{equation}
\begin{sbmatrix}{cc|cc}
\frac{1}{2} I + \KL_{b,b} & -\VL_{b,b} & & \\
\frac{1}{2} I - \KL_{b,b} & +\epsilon_{I,\mathit{II}} \VL_{b,b} & +\KL_{b,a} & -\VL_{b,a} \\ \hline
-\KL_{a,b} & +\epsilon_{I,\mathit{II}} \VL_{a,b} & \frac{1}{2} I + \KL_{a,a} & - \VL_{a,a} \\
& & \frac{1}{2} I - \KY_{a,a} & + \VY_{a,a}
\end{sbmatrix}
\begin{sbmatrix}{c}
\phi_b \\ \normderiv{\phi_b} \\ \phi_a \\ \normderiv{\phi_a}
\end{sbmatrix}
=
\begin{sbmatrix}{c}
\sum_i \frac{q_i}{\epsilon_I} G^L \\ 0 \\ 0 \\ 0
\end{sbmatrix}.
\label{eq:twoboundaryshorthand}
\end{equation}
Here, we have introduced a short-hand operator notation in which $I$
denotes the identity operator, $V$ denotes a single-layer potential
operator (the second term on the right-hand side of
Eq.~\ref{eq:Green-basic}) and $K$ denotes a double-layer potential
(the first term on the right-hand side of Eq.~\ref{eq:Green-basic});
the superscripts $L$ and $Y$ denote the Laplace or linearized
Poisson--Boltzmann (Yukawa) Green's function; and the subscript pair
$b,a$ denotes the ``source'' surface ($a$) and the ``destination''
surface ($b$).  The identity-operator terms arise from singularities
in the double-layer potential.

\subsection{Nonlocal-Response Electrostatic Model}
Figure~\ref{fig:problems}(b) is an illustration of the
nonlocal-response model.  As in the local-response problem, we assume
a spherical solute of radius $b$, centered at the origin, with $Q$
discrete point charges as the solute charge distribution $\rho(\bfr)$.
We denote the one spherical boundary in the problem, which separates
the protein and solvent, by $b$, and remind the reader that in this
problem we are only treating a single boundary.  Inside the protein,
the total electrostatic potential $\varphi_{\mathrm{I}}(\bfr)$ again
obeys the familiar local-response dielectric theory with dielectric
constant $\epsilon_{\mathrm{protein}}$:
\begin{align}
\mathbf{E}_{\mathrm{I}}&=-\nabla \varphi_{\mathrm{protein}},\\
  \mathbf{D}_{\mathrm{I}}(\bfr) &= \epsilon_{\mathrm{protein}} \epsilon_0 \mathbf{E}_{\mathrm{I}}(\bfr)\\
  \nabla \cdot \mathbf{D}_{\mathrm{I}}(\bfr) &= \rho(r).
\end{align}
We denote the Coulomb potential due to the fixed protein charges as
\begin{equation}
\varphi_{\mathrm{mol}} = \sum_{k=1}^Q \frac{q_k}{\epsilon_\mathrm{protein} |\bfr-\bfr_k|}
\end{equation}
and the reaction potential due to the difference between the protein
and solvent dielectric properties by $\varphi_{\mathrm{reac}}$, the
total electrostatic potential is 
\begin{equation}
  \varphi_{\mathrm{I}}(\bfr) = \varphi_{\mathrm{mol}}(\bfr) + \varphi_{\mathrm{reac}}(\bfr).
\end{equation}
In this nonlocal problem, we have a pure water solvent (no mobile
ions) in which the displacement and electric fields are related
\textit{nonlocally} by a convolution with a dielectric function of the
form $\mathcal{E}(\bfr,\bfrprime) = \epsilon(|\bfr-\bfrprime|)$ so
that
\begin{align}
  \mathbf{D}_{\mathrm{II}}(\bfr) &= \epsilon_0 \int_{\mathrm{II}} \mathcal{E}(\bfr,\bfrprime) \mathbf{E}_{\mathrm{II}}(\bfrprime) d^3 \bfrprime\label{eq:D-E-convolution}\\
  \nabla \cdot \mathbf{D}_{\mathrm{II}}(\bfr) &= 0,\label{eq:Gauss-nonlocal}
  \end{align}
and $\epsilon(|\bfr-\bfrprime|)$ is the Lorentz nonlocal function
\begin{equation}
\mathcal{E}(\bfr,\bfrprime)=\epsilon_\infty \delta(\bfr-\bfrprime) +
\frac{\epsilon_{\mathrm{water}}-\epsilon_\infty}{\lambda^2}\frac{\exp(-|\bfr-\bfrprime|/\lambda)}{4
  \pi |\bfr-\bfrprime|},\label{eq:Lorentzian-dielectric-function}
  \end{equation}
where $\epsilon_{\mathrm{water}}$ is the bulk solvent dielectric
constant (80 in the present work), $\epsilon_\infty$ is the
short-range dielectric constant, here taken to be the optical
dielectric constant 1.8, and $\lambda$ is an effective parameter that
reflects the length scale associated with correlations between solvent
molecules.  At the solute--solvent interface $b$, the usual Maxwell
boundary conditions Eqs.~\ref{eq:potential-bc}
and~\ref{eq:displacement-bc} apply.  By Eqs.~\ref{eq:D-E-convolution}
and~\ref{eq:Gauss-nonlocal}, the potential in the solvent must obey
not the familiar Laplace equation but instead the integrodifferential
equation
\begin{equation}
\nabla \cdot \int_{\mathrm{II}} \mathcal{E}(\bfr,\bfrprime) \nabla\varphi_{\mathrm{II}}(\bfrprime) d^3 \bfrprime = 0,
\end{equation}
the solution of which requires substantial calculation even for simple
cases such as a sphere with central
charge~\cite{Basilevsky96,Basilevsky98,Hildebrandt04,Hildebrandt05,Weggler_thesis}
or a charge approaching a planar
half-space~\cite{Hildebrandt04,Rubinstein04,Rubinstein07,Rubinstein10}.

Hildebrandt \textit{et al.} recently reformulated this nonlocal model
as a system of coupled but purely local partial differential equations
(PDEs)~\cite{Hildebrandt04}.  Similar simplification strategies have
been demonstrated for modeling dispersive electromagnetic
media~\cite{Ochs98} and plasticity~\cite{Engelen03}.  Essentially, for
a nonlocal relationship that takes the form of a Green's function for
a known PDE, one may be able to introduce a new unknown potential
whose gradient is the vector field resulting from the convolution
(here $\mathbf{D}_\mathrm{II}$).  Enforcing the original conservation
law (here, $\nabla \cdot \mathbf{D} = 0$) leads to an additional
Laplace equation and then the original unknown interest and the
additional unknown are coupled.  For the Lorentzian model, the
nonlocality resides in the second term of
Eq.~\ref{eq:Lorentzian-dielectric-function}, which is merely the
Green's function of the Yukawa equation $\nabla^2 u(\bfr) = \lambda^2
u(\bfr)$.  Here, by introducing the auxiliary displacement potential
$\psi_\mathrm{II}$, one may write the coupled PDE system as
\begin{align}
  \nabla^2 \varphi_\mathrm{I}(\mathbf{r}) = &-\rho(\mathbf{r})\\
  \nabla^2 \psi_\mathrm{II}(\mathbf{r}) = &0\\
  \left(\nabla^2 - \frac{1}{\Lambda^2}\right) \varphi_\mathrm{II}(\mathbf{r}) = &-\frac{1}{\lambda^2} \psi_\mathrm{II}(\mathbf{r})
  \end{align}
with $\Lambda = \lambda\sqrt{\epsilon_\infty /
  \epsilon_{\Sigma}}$. The exact displacement boundary condition
(Eq.~\ref{eq:displacement-bc}) is nonlocal and slow to compute, and so
Hildebrandt~\cite{Hildebrandt04} proposed the approximate boundary
conditions
\begin{align}
  \varphi_\mathrm{I}(\mathbf{r}_b^-) = & \varphi_\mathrm{II}(\mathbf{r}_b^+)\\
  \epsilon_0 \epsilon_\mathrm{protein} \frac{\partial}{\partial n} \varphi_\mathrm{I}(\mathbf{r}_b^-) = & \frac{\partial}{\partial n} \psi_\mathrm{II}(\mathbf{r}_b^+)\\
  \frac{\partial}{\partial n}\psi_\mathrm{II}(\mathbf{r}_b^+) = &\epsilon_0\epsilon_\infty \frac{\partial}{\partial n} \varphi_\mathrm{II}(\mathbf{r}_b^+).
\end{align}
Different choices for boundary conditions are analyzed in more detail
elsewhere, with model calculations suggesting that the impact on many
calculations should be small compared to the overall differences
between local and nonlocal models~\cite{Weggler_thesis}.

For numerical scaling, it is useful to change variables by introducing
the substitution
\begin{equation}
  \Psi = \frac{1}{\epsilon_\infty}\left(\frac{1}{\epsilon_0} \psi_\mathrm{II} - \epsilon_\mathrm{protein} \varphi_\mathrm{mol}\right),
  \end{equation}
as discussed extensively elsewhere~\cite{Hildebrandt05}.  Then,
defining
\begin{equation}
  b = -\left(\frac{1}{2}-\KYL+\frac{\epsilon_{\mathrm{protein}}}{\epsilon_{\mathrm{solvent}}}\KDR_\Lambda\right) \varphi_\mathrm{mol} - \left(\frac{\epsilon_{\mathrm{protein}}}{\epsilon_\infty}\VYL - \frac{\epsilon_{\mathrm{protein}}}{\epsilon_{\mathrm{solvent}}}\VDR_\Lambda\right)\normderiv{\varphi_\mathrm{mol}},\label{eq:b_rhs}
  \end{equation}
the complete BIE system is 
\begin{equation}
\begin{sbmatrix}{ccc}
\frac{1}{2}-\KYL & -\frac{\epsilon_{\mathrm{protein}}}{\epsilon_\infty}\VYL-\frac{\epsilon_{\mathrm{protein}}}{\epsilon_{\mathrm{solvent}}}\VDR_\Lambda & \frac{\epsilon_\infty}{\epsilon_{\mathrm{solvent}}}\KDR_\Lambda\\
\frac{1}{2}+\KL & -\VL & \\
& \frac{\epsilon_{\mathrm{protein}}}{\epsilon_\infty} \VL & \frac{1}{2}-\KL
  \end{sbmatrix}
\begin{sbmatrix}{c}
\varphi_{\mathrm{II}} \\ \normderiv{\varphi_{\mathrm{II}}} \\ \Psi
  \end{sbmatrix}
=
\begin{sbmatrix}{c} b \\ 0 \\ 0 \end{sbmatrix},\label{eq:full-formulation}
\end{equation}
where $V^{DR}_\Lambda = V^Y_\Lambda - V^L$ and similarly
$K^{DR}_\Lambda = K^Y_\Lambda - K^L$.  We omit the lengthy derivation
and refer interested readers to Hildebrandt~\cite{Hildebrandt05}.

A point of great importance for fast numerical solution of
Eq.~\ref{eq:full-formulation} is that each non-zero block is a linear
combination of the same boundary integral operators as are needed to
solve Eq.~\ref{eq:twoboundaryshorthand}.  As a result, the same fast
BEM solvers used for local electrostatics in the LPBE model (e.g.,
fast multipole methods~\cite{Lu06}, pre-corrected FFT~\cite{Kuo02},
and the FFTSVD algorithm~\cite{Altman06,Altman09}) can be adapted
easily to solve nonlocal electrostatics models~\cite{Bardhan11_DAC}.
Fast solvers allow the discretized linear system, which is dense in
the sense that the number of non-zero entries grows quadratically with
the number of unknowns, to be solved in linear or near-linear time.

\subsection{Eigenfunction Expansions of Boundary-Integral Operators on Spheres}
All of the boundary-integral operators of
Eqs.~\ref{eq:twoboundaryshorthand} and~\ref{eq:full-formulation} are
diagonalized by the surface spherical harmonics~\cite{Hsiao94}.
Consequently, the boundary integrals of the form $\int
F(\mathbf{r},\mathbf{r}') u(\mathbf{r}') dA'$ can be re-written as
\begin{equation}
\int F(\mathbf{r},\mathbf{r}') u(\mathbf{r}') dA'
= \sum_{n=0}^{\infty} \sum_{m=-n}^{+n} Y^n_m(\theta,\phi) \lambda^F_{nm} \int Y^{n,*}_m(\theta',\phi') u(\theta',\phi') dA'\label{eq:operator-eigendecomposition}
\end{equation}
where the $(\theta,\phi)$ are the angular coordinates for
$\mathbf{r}$, $Y^n_m(\theta,\phi)$ are the orthonormal surface
harmonics, and $\lambda^{F}_{nm}$ is the eigenvalue for the $n,m$
mode.  Note that Eq.~\ref{eq:operator-eigendecomposition} represents a
slight abuse of notation, in that the radii of the ``source'' and
``destination'' spheres are included only implicitly in the actual
eigenvalues.  Also, in this work, the eigenvalues of the relevant
operators are independent of $m$, so we omit the second subscript in
the remainder of the text.

For a sphere of radius $R$, the eigenvalues of the four
``self-to-self'' operators $\VL$, $\KL$, $\VY$, and $\KY$ are
\begin{align}
  \lambda_n^{\VL} &= \frac{R}{2 n + 1}\\
  \lambda_n^{\KL} &= -\frac{1}{2 (2 n + 1)}\\
  \lambda_n^{\VY} &= i (i\kappa) R^2 j_n(i \kappa R) h_n^{(1)}(i \kappa R)\\
  \lambda_n^{\KY} &= i (i\kappa)^2 R^2/2 \left(j_n(i \kappa R) h_n^{(1)}(i \kappa R)\right)^\prime\label{eq:lambda-KYL}
\end{align}
where $i=\sqrt{-1}$, $j_n(x)$ and $h_n^{(1)}(x)$ denote the spherical
Bessel function and spherical Hankel function of the first kind,
respectively, and the prime notation in Eq.~\ref{eq:lambda-KYL}
denotes differentiation with respect to the argument.

The Kirkwood problem also involves four Laplace boundary-integral
operators that map between concentric spheres.  We demonstrate in the
Appendix that the eigenvalues of these operators are
\begin{align}
  \lambda_n^{\VL_{a,b}} &= \left(\frac{b}{a}\right)^{n+1} \frac{b}{2 n + 1}\\
  \lambda_n^{\KL_{a,b}} &= \begin{cases} 0,& n = 0\\
      -2 n \left(\frac{b}{a}\right)^{n+1} \frac{-1}{2 (2 n + 1)}, & n > 0
      \end{cases}\\
  \lambda_n^{\VL_{b,a}} &= \left(\frac{a}{b}\right)^n \frac{a}{2 n + 1} \\
  \lambda_n^{\KL_{b,a}} &=  \begin{cases} 1, & n = 0\\
    2(n+1) \left(\frac{a}{b}\right)^n \frac{-1}{2 (2 n + 1)}, & n > 0.
  \end{cases}
  \end{align}

\section{The Boundary-Integral-Equation + Eigenfunction Approach}\label{sec:bie-eigenfunctions}
\subsection{Application to the Kirkwood Problem}
To simplify the coupled boundary-integral equations, we introduce the
spherical-harmonic projection operator $Y^*$, which maps a function
defined on a sphere (i.e. in angular coordinates) into the expansion
coefficients in the basis of surface spherical harmonics, which is
complete and orthonormal.  Similarly, the operator $Y$ maps a vector
of expansion coefficients in the basis of surface harmonics to a
function on the sphere.

The non-zero blocks of the matrix in Eq.~\ref{eq:twoboundaryshorthand}
can be simultaneously diagonalized as
\begin{eqnarray}
\begin{sbmatrix}{cc|cc}
  \frac{1}{2} + D^{(1)} & -D^{(2)} & & \\
  \frac{1}{2} - D^{(1)} & +\epsilon_{I,II} D^{(2)} & +D^{(3)} & -D^{(4)}\\ \hline
  -D^{(5)} & +\epsilon_{I,II} D^{(6)} & \frac{1}{2} + D^{(7)} & -D^{(8)}\\
  & & \frac{1}{2} -D^{(9)} & +D^{(10)}
  \end{sbmatrix}\nonumber\\
=
\begin{sbmatrix}{cccc}
    Y^* & 0 & 0 & 0\\
    0 & Y^* & 0 & 0\\
    0 & 0 & Y^* & 0\\
    0 & 0 & 0 & Y^*
  \end{sbmatrix}
\begin{sbmatrix}{cc|cc}
\frac{1}{2} I + \KL_{b,b} & -\VL_{b,b} & & \\
\frac{1}{2} I - \KL_{b,b} & +\epsilon_{I,\mathit{II}} \VL_{b,b} & +\KL_{b,a} & -\VL_{b,a} \\ \hline
-\KL_{a,b} & +\epsilon_{I,\mathit{II}} \VL_{a,b} & \frac{1}{2} I + \KL_{a,a} & - \VL_{a,a} \\
& & \frac{1}{2} I - \KY_{a,a} & + \VY_{a,a}
\end{sbmatrix}
\begin{sbmatrix}{cccc}
    Y & 0 & 0 & 0\\
    0 & Y & 0 & 0\\
    0 & 0 & Y & 0\\
    0 & 0 & 0 & Y
  \end{sbmatrix},
\end{eqnarray}
with $D^{(1)}_{ii} = \lambda_{n(i)}^{\KL}|_{R=b}$,
$D^{(2)}_{ii} = \lambda_{n(i)}^{\VL}|_{R=b}$,
  $D^{(3)}_{ii} = \lambda_{n(i)}^{\KL_{b,a}}$,
  $D^{(4)}_{ii} = \lambda_{n(i)}^{\VL_{b,a}}$,
  $D^{(5)}_{ii} = \lambda_{n(i)}^{\KL_{a,b}}$,
  $D^{(6)}_{ii} = \lambda_{n(i)}^{\VL_{a,b}}$,
  $D^{(7)}_{ii} = \lambda_{n(i)}^{\KL}|_{R=a}$,
  $D^{(8)}_{ii} = \lambda_{n(i)}^{\VL}|_{R=a}$,
  $D^{(9)}_{ii} = \lambda_{n(i)}^{\KY}|_{R=a}$,
  and $D^{(10)}_{ii} = \lambda_{n(i)}^{\VY}|_{R=a}$,
where $n(i)$ denotes the degree associated with the $i$th eigenmode.
Expanded in the surface harmonics, the unknowns of
Eq.~\ref{eq:twoboundaryshorthand} are written
\begin{equation}
  \begin{sbmatrix}{c}
\tilde{\phi_b} \\ \tilde{\normderiv{\phi_b}} \\ \tilde{\phi_a} \\ \tilde{\normderiv{\phi_a}}
    \end{sbmatrix}
  =
  \begin{sbmatrix}{cccc}
    Y^* & 0 & 0 & 0\\
    0 & Y^* & 0 & 0\\
    0 & 0 & Y^* & 0\\
    0 & 0 & 0 & Y^*
    \end{sbmatrix}
  \begin{sbmatrix}{c}
\phi_b \\ \normderiv{\phi_b} \\ \phi_a \\ \normderiv{\phi_a},
    \end{sbmatrix}
  \end{equation}
and projecting the right-hand side similarly, we obtain the
surface-harmonic analogue to Kirkwood's result:
\begin{equation}
\begin{sbmatrix}{cc|cc}
  \frac{1}{2} + D^{(1)} & -D^{(2)} & & \\
  \frac{1}{2} - D^{(1)} & +\epsilon_{I,II} D^{(2)} & +D^{(3)} & -D^{(4)}\\ \hline
  -D^{(5)} & +\epsilon_{I,II} D^{(6)} & \frac{1}{2} + D^{(7)} & -D^{(8)}\\
  & & \frac{1}{2} -D^{(9)} & +D^{(10)}
  \end{sbmatrix}
  \begin{sbmatrix}{c}
\tilde{\phi_b} \\ \tilde{\normderiv{\phi_b}} \\ \tilde{\phi_a} \\ \tilde{\normderiv{\phi_a}}
  \end{sbmatrix}
  =
  \begin{sbmatrix}{c}
   Y^* \varphi_{\mathrm{mol}} \\ 0 \\ 0 \\ 0
    \end{sbmatrix}.\label{eq:modal-Kirkwood}
  \end{equation}
Note that this representation does \textit{not} diagonalize the entire
operator, but does decompose the reaction potential in the protein into
the individual harmonics.

An algorithm to solve the Kirkwood problem using the BIE/eigenfunction
approach is therefore structured as follows.  For each mode $i$ to be
solved (up to a desired order), one first computes the projection of
the solute charge distribution onto the $i$th solid spherical harmonic
(i.e. one computes the appropriate multipole expansion coefficient).  Then
one calculates the $i$th eigenvalues for the boundary integral operators
to set up a linear system of equations with four unknowns, and solves for
the $i$th expansion coefficient of the reaction potential.  The reaction
potentials at all desired locations is then easily computed.

\subsection{Application to Nonlocal Electrostatics}
We now derive our main result---the exact analytical solution of
nonlocal electrostatics for a spherical solute.  The 3-by-3 block
operator of Eq.~\ref{eq:full-formulation} can be decomposed as
\begin{equation}
  \begin{sbmatrix}{ccc}
    Y & & \\ & Y & \\ & & Y
  \end{sbmatrix}
  \begin{sbmatrix}{ccc}
    D^{(1)} & D^{(2)} & D^{(3)} \\ D^{(4)} & D^{(5)} & \\ & D^{(6)} & D^{(7)}
  \end{sbmatrix}
  \begin{sbmatrix}{ccc}
    Y^* & & \\ & Y^* & \\ & & Y^*
  \end{sbmatrix},\label{eq:diagonal-block}
\end{equation}
where again $Y^*$ projects from a distribution on the sphere surface
into an expansion in surface spherical harmonics, $Y$ represents the
harmonics themselves, and the matrices $D^{(k)}$ are all diagonal.
The entries of the $D^{(k)}$ matrices are simply the appropriate
scaled sum of the operator eigenvalues: e.g., $D^{(1)}_{ii} =
\frac{1}{2}-\lambda^{K_\Lambda}_{n(i)}$, where $n$ represents the
degree associated with the $i$th eigenmode.  Projecting both sides of
Eq.~\ref{eq:full-formulation}, one obtains
\begin{equation}
  \begin{sbmatrix}{ccc}
    D^{(1)} & D^{(2)} & D^{(3)} \\ D^{(4)} & D^{(5)} & \\ & D^{(6)} & D^{(7)}
  \end{sbmatrix}
  \begin{sbmatrix}{c}
    \tilde{\varphi}_{\mathrm{II}} \\
    \normderiv{\tilde{\varphi}_{\mathrm{II}}} \\
    \tilde{\Psi}
    \end{sbmatrix}
  =
  \begin{sbmatrix}{c}
    \tilde{b} \\ 0 \\ 0
    \end{sbmatrix},
  \end{equation}
The $i$th entry of the projected form of Eq.~\ref{eq:b_rhs} is therefore
\begin{equation}
\tilde{b}_{i} = -\left(\frac{1}{2}-\lambda^{\KYL}_{n(i)}+\frac{\epsilon_{\mathrm{protein}}}{\epsilon_{\mathrm{solvent}}}\lambda^{\KDR_\Lambda}_{n(i)}\right)\tilde{\varphi}_\mathrm{mol}-\left(\frac{\epsilon_{\mathrm{protein}}}{\epsilon_{\mathrm{\infty}}}\lambda^{\VYL}_{n(i)}-\frac{\epsilon_{\mathrm{protein}}}{\epsilon_{\mathrm{solvent}}}\lambda^{\VDR_\Lambda}_{n(i)}\right)\normderiv{\tilde{\varphi}_\mathrm{mol}}.
\end{equation}
Again, solving analytically for each coefficient
$\tilde{\varphi}_{n(i)}$ independently provides the desired expansion
(in surface harmonics) of the potential at the protein-water boundary.
These coefficients are readily converted to the solid harmonics to
obtain the potential inside the sphere.  The analytical nonlocal model
has been implemented in MATLAB and is available as Supplemental
Information~\cite{analytical-nonlocal-SI,analytical-nonlocal-bitbucket}.

\begin{figure}[ht!]
  \centering \resizebox{6.0in}{!}{\includegraphics{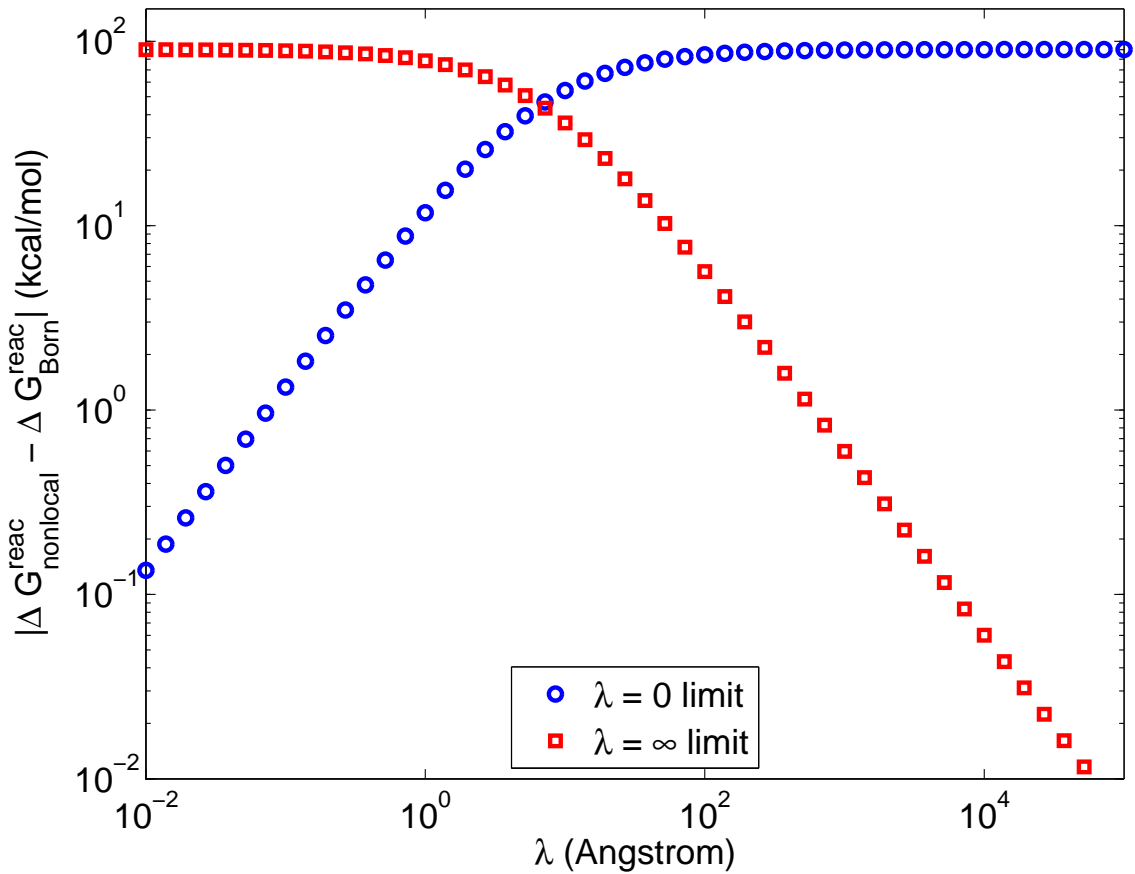}}
  \caption{The analytically computed solvation free energy for a
    sphere with central charge (Born ion) converges to the correct
    local-response limits as the nonlocal length-scale parameter
    $\lambda$ approaches 0
    or~$\infty$.}\protect\label{fig:convergence-to-local} \end{figure}

\begin{figure}[ht!]
  \centering
  \resizebox{6.0in}{!}{\includegraphics{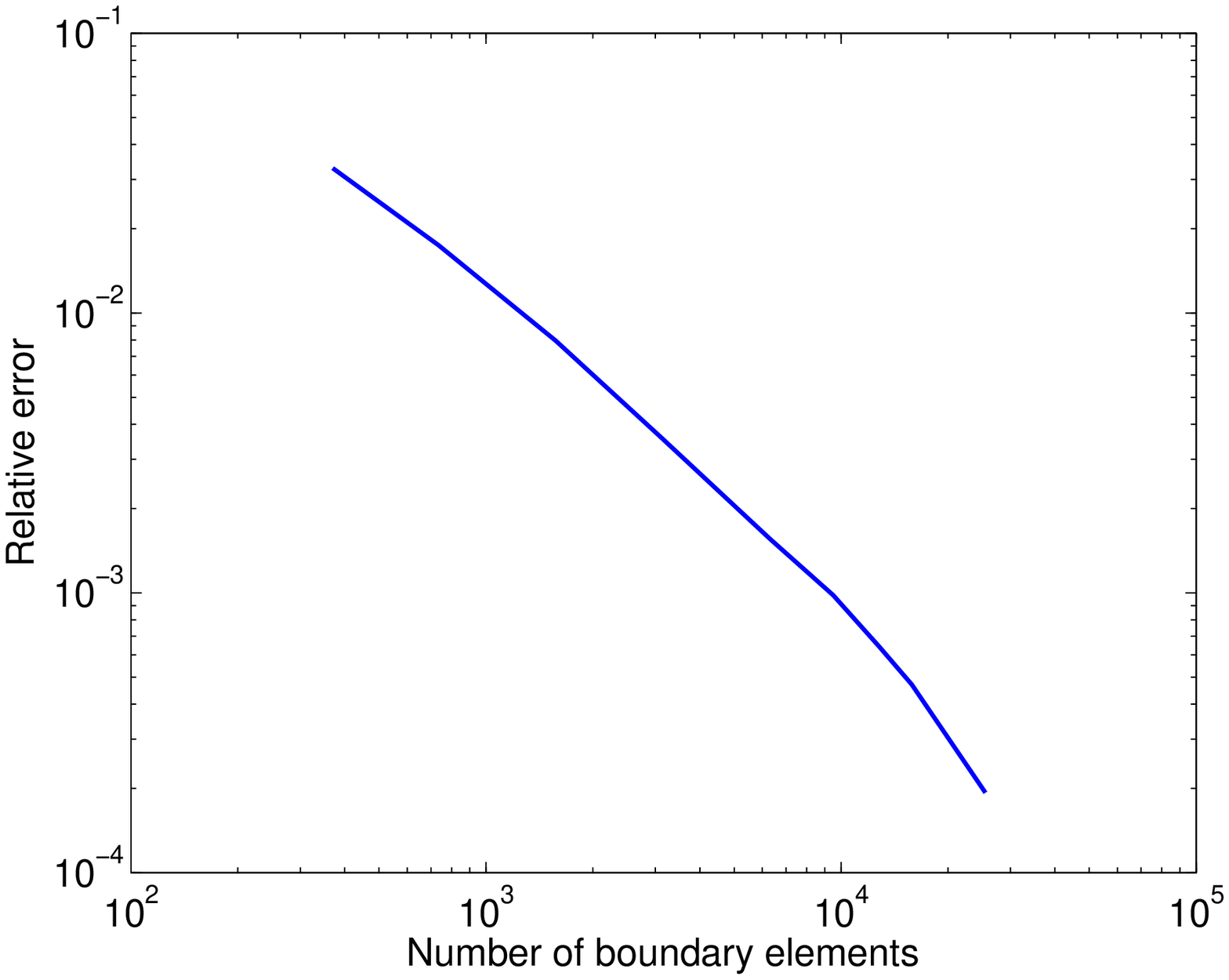}}
  \caption{Relative error for numerical simulations of the nonlocal
  model using BEM, as a function of the number of unknowns in the
  discretized problem, for a 8-\AA-radius sphere with a single $+1e$
  charge situated 2~\AA~from the sphere
  surface.}\protect\label{fig:convergence-sph8-q-at-6} \end{figure}

It may be verified that in the limits $\lambda \to 0$ and $\lambda \to
\infty$, the analytical solution converges to the appropriate
local-response models; see Figure~\ref{fig:convergence-to-local} for
the example of a sphere with a single central charge, which is known
as the Born ion.  As a more challenging validation, we have used the
nlFFTSVD fast BEM solver~\cite{Bardhan11_DAC} to compute the solvation
free energy of a single $+1e$ charge situated at (0, 0, 6~\AA) inside
a sphere of radius 8~\AA~centered at the origin, and plotted the
convergence of these results to the solvation free energy computed
analytically (Figure~\ref{fig:convergence-sph8-q-at-6}).  This test
case is challenging because it lacks the spherical symmetry of the
Born-ion test case, and in fact BEM simulations require finer
discretization for charges close to the surface~\cite{Juffer91}.

The required spherical Bessel and Hankel functions have been computed
using the algorithm proposed by Cai~\cite{Cai11}, and their
derivatives were calculated using well-known recurrence
relations~\cite{VanBladel}.  Using numerically stable implementations
of the Bessel functions and their derivatives is of utmost importance.
For large sphere radii and charges approaching the surface, large
cancellations in na\"ive implementations of the projections causes the
observed value of the solvation energy to diverge as the order of the
calculation is increased.  Additionally, very large and very small
values of $\lambda$ are problematic for the calculation of the
Yukawa-operator eigenvalues, and suggest that further research into
their accurate computation is warranted.

\section{Results}\label{sec:results}
\begin{figure}[ht!]
  \centering
  \resizebox{6.0in}{!}{\includegraphics{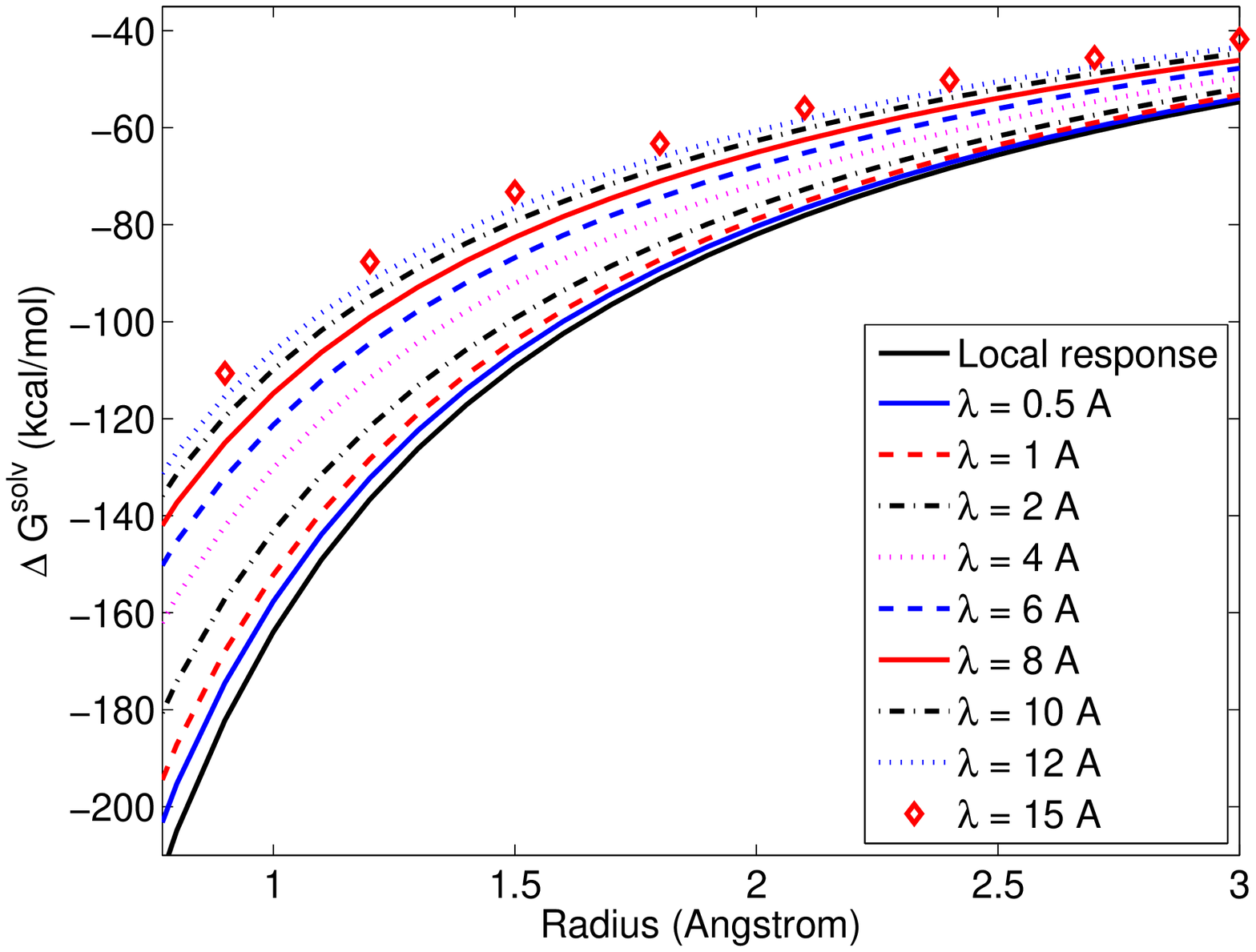}}
  \caption{Dependence of the electrostatic solvation free energy for a
    $+1e$ point charge centrally located in a sphere, as a function of
    the sphere radius and the nonlocal parameter $\lambda$.  Here
    $\epsilon_\mathrm{protein} = 1$, $\epsilon_\mathrm{water} = 80$,
    and $\epsilon_\infty =
    1.8$.}\protect\label{fig:expt-ions-param-sweep}
\end{figure}

Hildebrandt \textit{et al.} have suggested that $\lambda \approx
15-24$~\AA~provided an excellent fit to experimental data for
monatomic cations~\cite{Hildebrandt04,Hildebrandt05}.  However, other
factors, especially nonlinearities such as dielectric
saturation~\cite{Gong08}, may play important roles in ion solvation
and charge burial, it is important to understand the dependence on
$\lambda$.  Figure~\ref{fig:expt-ions-param-sweep} contains plots of
nonlocal-model electrostatic solvation free energies for monovalent
cations of varying radii.  The nonlocal-model free energies are
clearly sensitive to $\lambda$ in the range 1 to 10~\AA, but much less
so outside that range.  Therefore, although ion solvation free
energies therefore provide a clear and intuitive demonstration of the
impact of nonlocal response, the insensitivity outside of $\lambda >
10$~\AA~suggests that that such data should be used with caution when
performing more detailed parameterization; in particular, we emphasize
that it is impossible to fit the ion radii as well as $\lambda$
simultaneously because the problem is underdetermined.  More extensive
calculations are needed to calibrate the new model, and are underway.

Analytical solutions for simple geometries also allow fast
determination of the reaction potential throughout the whole system.
More thorough visualizations of solvent response may offer new
insights into the empirical, seemingly application-specific
definitions of the protein dielectric constant~\cite{Sham98,Schutz01},
including for example why values of $\epsilon_\mathrm{protein}$ much
larger than experimental estimates~\cite{Gilson86} are often needed to
obtain accurate calculations of p$K_\mathrm{a}$ shifts in
proteins~\cite{Demchuk96}.  To illustrate the fundamental differences
between local and nonlocal theory, as well as the computational
advantage of having a fast analytical model for visualization, we plot
the reaction potentials for both simple and complicated charge
distributions as we vary key model parameters: the protein dielectric
constant in the local theory, and the effective length scale $\lambda$
in the nonlocal model.

\begin{figure}[ht!]
  \centering
  \resizebox{6.0in}{!}{\includegraphics{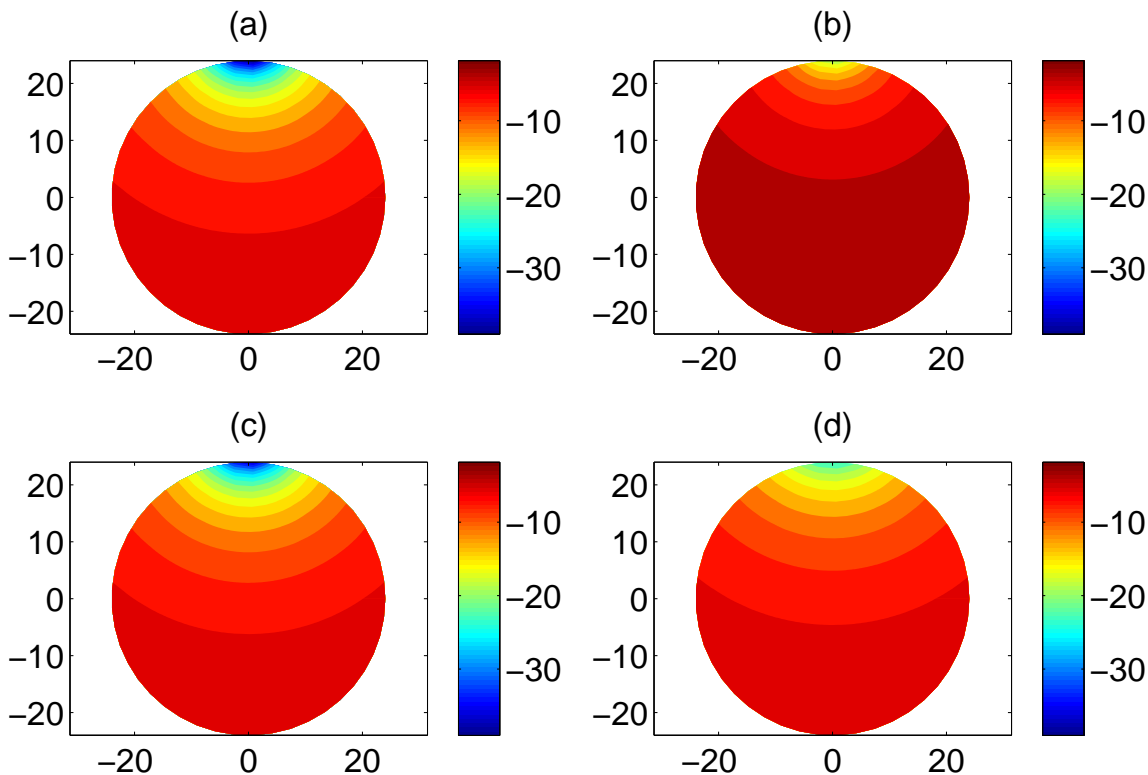}}
  \caption{Reaction potential (in kcal/mol/$e$) induced in a sphere of
    radius 24 \AA~by a single $+1e$ point charge situated 2 \AA~from
    the boundary, for different local and nonlocal models.  All
    potentials are plotted on the same color scale, and for all
    models, $\epsilon_\mathrm{water} = 80$. (a) Local-response model
    with $\epsilon_\mathrm{protein} = 2$; (b) local-response model
    with $\epsilon_\mathrm{protein} = 4$; (c) nonlocal-response model
    with $\epsilon_\mathrm{protein} = 2$ and $\lambda = 1$~\AA; (d)
    nonlocal-response model with $\epsilon_\mathrm{protein} = 2$ and
    $\lambda =
    10$~\AA.}\protect\label{fig:reac-pot-vary-epsilon-surface-charge}
\end{figure}

\begin{figure}[ht!]
  \centering
  \resizebox{6.0in}{!}{\includegraphics{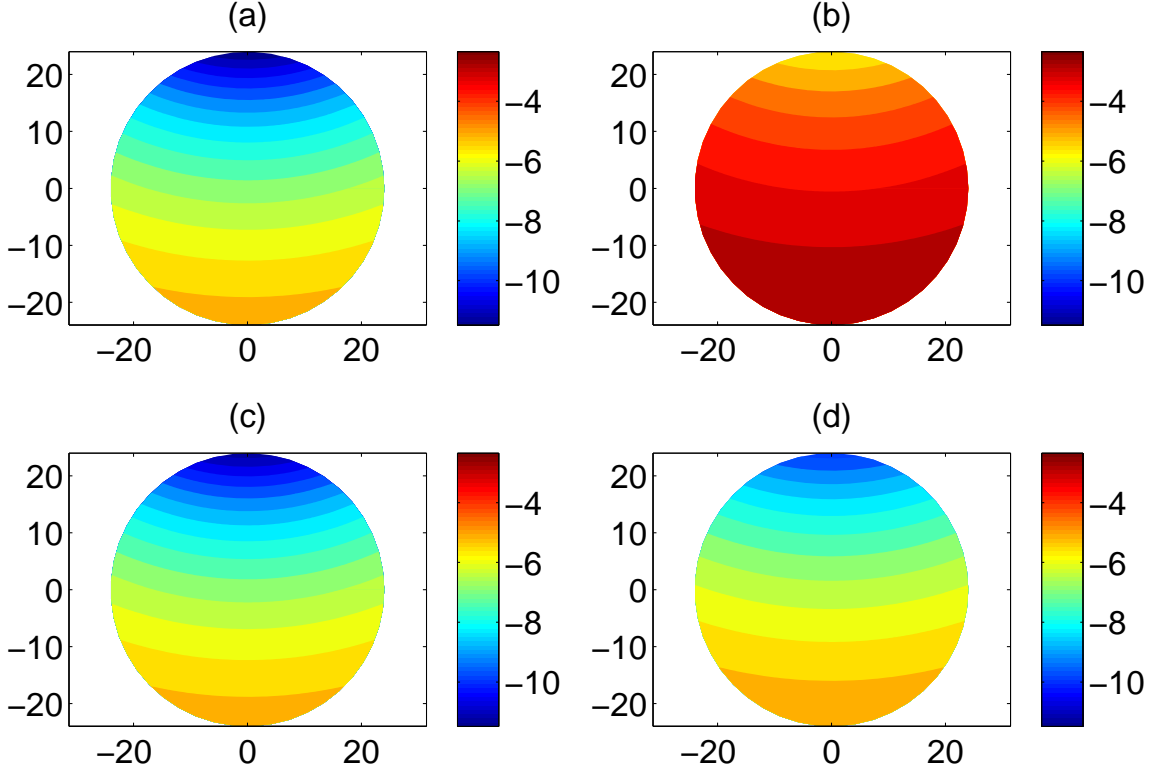}}
  \caption{Reaction potential (in kcal/mol/$e$) induced in a sphere of
    radius 24 \AA~by a single $+1e$ point charge buried 14 \AA~from
    the boundary, for different local and nonlocal models.  All model
    parameters are the same as the corresponding plots in
    Figure~\ref{fig:reac-pot-vary-epsilon-surface-charge}.
  }\protect\label{fig:reac-pot-vary-epsilon-buried-charge}
\end{figure}

Figure~\ref{fig:reac-pot-vary-epsilon-surface-charge} contains plots
of the reaction potential induced by a single $+1e$ charge in a
protein-sized sphere of radius 24~\AA, where the charge is situated
2~\AA~from the dielectric boundary.  The reaction potential for
local-response models is shown in (a) and (b), with
$\epsilon_\mathrm{protein} = 2$ in (a) and $\epsilon_\mathrm{protein}
= 4$ in (b).  Nonlocal-model results are plotted in (c) and (d); for
both nonlocal calculations, $\epsilon_\mathrm{protein} = 2$, with
$\lambda = 1$~\AA~in (c) and $\lambda = 10$~\AA~in (d).  For comparison,
all potentials are plotted according to the same color scale.
Adjusting $\epsilon_\mathrm{protein}$ from 2 to 4 in the local model
leads to a qualitative global shift in the reaction potential.  On the
other hand, nonlocal response presents relatively small overall
changes, even though $\lambda$ varies substantially.  For a single
$+1e$ charge buried deep within the protein at (0, 0, 10~\AA), the
reaction potential is smaller in magnitude, which means that the
qualitative shift for increased $\epsilon_\mathrm{protein}$ can be
seen more easily
(Figure~\ref{fig:reac-pot-vary-epsilon-buried-charge}).

\begin{figure}[ht!]
  \centering
  \resizebox{6.0in}{!}{\includegraphics{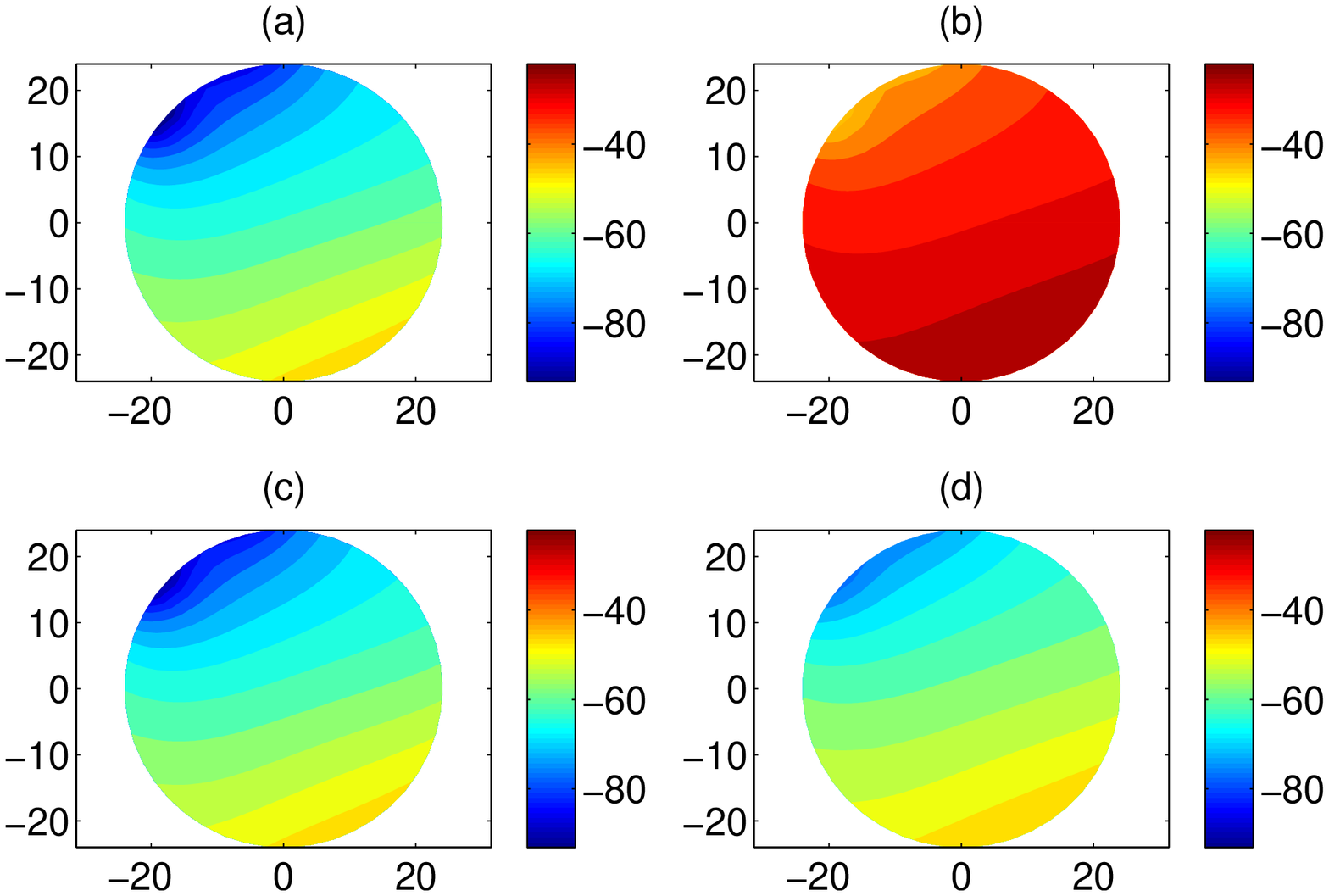}}
  \caption{Reaction potential (in kcal/mol/$e$) induced in a sphere of
    radius 24 \AA~by the charge distribution of the protein bovine
    pancreatic trypsin inhibitor (BPTI), for different local and
    nonlocal models.  All model parameters are the same as the
    corresponding plots in
    Figure~\ref{fig:reac-pot-vary-epsilon-surface-charge}.  See main
    text for full computational
    details.}\label{fig:reac-pot-vary-epsilon-bpti}
\end{figure}

These qualitative differences are meaningful for the types of
complicated charge distributions found in proteins as well.  To
illustrate this point, we use as an example the protein bovine
pancreatic trypsin inhibitor (BPTI).  We model the charge distribution
by taking the atomic coordinates from the Protein Data Bank (accession
code 3BTM~\cite{Helland99}) and assigning atomic charges using the
PARSE~\cite{Sitkoff94} force field.
Figure~\ref{fig:reac-pot-vary-epsilon-bpti} contains plots of the
resulting reaction potentials; the results for each subfigure are
computed using the same model and parameters as used for the
corresponding subfigure of
Figure~\ref{fig:reac-pot-vary-epsilon-surface-charge}.  Together,
these results suggest that future nonlocal studies should investigate
charge-charge interactions in more detail, especially contrasting the
fields induced by buried and surface-exposed charges.


We would like to emphasize the substantial difference in speed between
numerical and analytical methods.  For the BPTI test problem, a
low-resolution numerical simulation using the highly optimized,
linear-scaling boundary-element method (BEM) code
nlFFTSVD~\cite{Bardhan11_DAC}---one of the fastest numerical
implementations of the nonlocal model---requires approximately 12
minutes on a 2012 MacBook Air.  The unoptimized MATLAB implementation
of the analytical approach, in contrast, is more than 45 times faster,
requiring less than 17 seconds.

\section{Discussion}\label{sec:discussion}
The shortcomings of local electrostatics continue to motivate new
models, but often the practical complications of numerical simulation
slow their testing and improvement.  To accelerate studies of the promising
Lorentz nonlocal
model~\cite{Hildebrandt04,Hildebrandt07,Weggler10,Bardhan11_pka}, we
have derived the exact analytical solution for a spherical solute
containing an arbitrary charge distribution.  Our approach uses
Hildebrandt's boundary-integral equation (BIE) formulation
~\cite{Hildebrandt07} and the analytically known eigendecompositions
of the associated boundary-integral operators.  Calculations
demonstrate the method's correctness and that solvent screening of
charge-charge interactions are markedly different in nonlocal and
local theories, even when the protein dielectric constant is adjusted.
Fast analytical models enable rapid visualization of electrostatic
fields, and thus facilitates efficient exploration of the new model's
implications and qualitative differences from existing theories.

The BIE-eigenfunction strategy represents a novel alternative to
matching potential expansions and may be useful in other areas of
mathematical physics.  To illustrate the method's generality, we have
also derived the solution to the Kirkwood two-boundary problem for
local electrostatics, which has furnished many insightful physical
studies and model approximations even though proteins clearly take
shapes much more complex than spheres.  It should also be noted that a
similar algorithmic approach can simplify calculations involving
matched solid-harmonic expansions; that is, instead of tedious,
error-prone algebraic manipulations to obtain the desired expansion
coefficients, one could set up the small linear systems for each mode
and allow the computer to do the arithmetic.

The present work enables studies of the nonlocal model to be conducted
rapidly for simple model systems, obviating the need for more
complicated and slower numerical
calculations~\cite{Hildebrandt07,Weggler10,Bardhan11_DAC,Bardhan11_pka}.
To encourage further tests of nonlocal models, the Supplemental
Information~\cite{analytical-nonlocal-SI,analytical-nonlocal-bitbucket}
includes a MATLAB implementation of the analytical approach. As
described in earlier work on nonlocal electrostatics, boundary
conditions represent a subtle issue that warrants detailed
study~\cite{Hildebrandt05,Weggler10,Fasel08}, and fast calculations on
spheres will allow a simple way to test improvements.  Our results
also provide a useful way to test numerical simulations of nonlocal
electrostatics on nontrivial systems, e.g. models of finite-sized
solutes with complicated charge distributions.

Future work will address the development of fast analytical
approximations similar to recent Generalized-Born (GB)
models~\cite{Sigalov05} or BIE
approximations~\cite{Bardhan11_Knepley}.  Second-kind
boundary-integral formulations may offer substantial advantages for
such approximations~\cite{Liang97}, and Fasel et al. have recently
presented a purely second-kind formulation of the nonlocal
model~\cite{Fasel08}.  An extension of our approach to the Fasel
formulation is therefore of significant interest.  One extension to
the present work might be to account for the fact that many proteins
can be reasonably well modeled using ellipsoids (see, for a recent
example in electrostatic theory,~\cite{Sigalov06}).  It is possible
that one could use a similar approach to derive an analytical solution
for ellipsoidal geometries as well; the eigendecompositions of the
Laplace boundary-integral operators for ellipsoids are known, for
instance~\cite{Ahner86,Ahner96,Ritter95_spectrum}, though
corresponding results for the Yukawa integral operators do not appear
to have been published.  We also note that even for the sphere,
computing the eigenvalues of the Yukawa integral operators is
numerically challenging, and should motivate the development of
improved algorithms.  Recent work on computing the ellipsoidal
harmonics found similar
challenges~\cite{Bardhan12_Knepley_ellipsoidal}, and the present work
has uncovered a second compelling example of how molecular biophysics
poses novel challenges for more fundamental research in applied
mathematics and numerical analysis.


\section*{Appendix: Eigenvalues of the Laplace boundary-integral operators for concentric spheres}

We first address the single- and double-layer operators that map from
the inner sphere (radius $b$) to the outer (radius $a$).  For the
single-layer operator, let us expand a surface potential on the inner
sphere in surface harmonics, i.e. 
\begin{equation}
  \psi_{S_b} = \sum_{n,m} S^b_{nm} Y^m_n(\theta,\phi).
\end{equation}
and also expand the potential field in the region outside that sphere
\begin{equation}\label{eq:outerVolumeField}
  \psi = \sum_{n,m} V_{nm} r^{-(n+1)} Y^m_n(\theta,\phi).
\end{equation}
These two fields must agree on the surface $r = b$, and by
orthogonality of the $Y^m_n$ functions, we have
\begin{equation}\label{eq:innerExp}
  V_{nm} = S^b_{nm} b^{n+1}.
\end{equation}
A similar surface expansion holds for the fields on the outer concentric sphere
\begin{equation}
  \psi_{S_a} = \sum_{n,m} S^a_{nm} Y^m_n(\theta,\phi),
\end{equation}
which may be matched to Eq.~\ref{eq:outerVolumeField} to give
\begin{equation}\label{eq:outerExp}
  V_{nm} = a^{n+1} S^a_{nm}.
\end{equation}
Combining Eq.~\ref{eq:innerExp} with Eq.~\ref{eq:outerExp}, we have
\begin{equation}
  S^a_{nm} = \frac{b^{n+1}}{a^{n+1}} S^b_{nm}
\end{equation}
Because the eigenvalue for the single-layer Laplace surface operator
on the inner surface is $b / (2n + 1)$, we finally have that
\begin{equation}
  \lambda^{V^L_{a,b}}_n = \left(\frac{b}{a}\right)^{n+1} \frac{b}{2n + 1}.\label{eq:eigenvalues-concentric-single}
\end{equation}

We derive the double-layer potential operators using an alternative
approach based on Green's theorem.  Consider again the expansion in
spherical harmonics of the potential outside $b$ from
Eq.~\ref{eq:outerVolumeField}, so that the radial component of the
electric field is
\begin{equation}
\frac{\partial \psi}{\partial r} = \sum_{n,m} V_{nm} (-(n+1)) r^{-(n+2)} Y^m_n(\theta,\phi),
  \end{equation}
so the normal derivative of the potential at the inner surface $b$ is
\begin{equation}
\frac{\partial \psi}{\partial r}|_{r=b} = \sum_{n,m} V_{nm} (-(n+1)) b^{-(n+2)} Y^m_n(\theta,\phi).
  \end{equation}
Green's theorem allows us to write the potential at any point
$\mathbf{r}$ with $r=a > b$ as
\begin{equation}
  \psi(\mathbf{r}) = +\int_b \frac{\partial G(\mathbf{r},\mathbf{r}')}{\partial n} \psi(\mathbf{r}') dA' - \int_b G(\mathbf{r},\mathbf{r}') \frac{\partial \psi(\mathbf{r}')}{\partial n} dA'\label{eq:Green1}.
  \end{equation}
Using Eq.~\ref{eq:operator-eigendecomposition} and again relying on
the orthogonality of the harmonics, we obtain
\begin{equation}
  a^{-(n+1)}= \lambda_{n}^{K^L_{a,b}}b^{-(n+1)} - \lambda_{n}^{V^{L}_{a,b}} (-(n+1)) b^{-(n+2)};
  \end{equation}
substituting the known $\lambda_{n}^{V^L_{a,b}}$ from
Eq.~\ref{eq:eigenvalues-concentric-single} gives
\begin{equation}
  a^{-(n+1)} = \lambda_{n}^{\KL} b^{-(n+1)} + \frac{n+1}{2 n + 1} a^{-(n+1)}
  \end{equation}
and finally
\begin{equation}
\lambda^{\KL}_{n} = \frac{n}{2 n + 1}\left(\frac{b}{a}\right)^{n+1}.
  \end{equation}
This result may be checked in the limit as $a \rightarrow b$, where
Eq.~\ref{eq:Green1} becomes
\begin{equation}
    \psi(\mathbf{r}) = \frac{1}{2} \psi(\mathbf{r}) +\int
    \frac{\partial G(\mathbf{r},\mathbf{r}')}{\partial n} \psi(\mathbf{r}') dA' - \int G(\mathbf{r},\mathbf{r}') \frac{\partial\psi(\mathbf{r}')}{\partial n} dA'.
  \end{equation}
Analogous manipulations lead to the relation
\begin{equation}
  a^{-(n+1)} = \frac{1}{2} a^{-(n+1)} + \lambda_{nm}^K b^{-(n+1)} + \frac{n+1}{2 n + 1} a^{-(n+1)}
  \end{equation}
and thus we recover the self-surface result that $\lambda_{n}^{\KL} =
\frac{-1}{2(2n+1)}$.
The eigenvalues for the operators that map from the outer sphere to
the inner one are obtained in very similar fashion using interior
harmonics.  For example, 
\begin{equation}\label{eq:SecondOuterVolumeField}
  \psi = \sum_{n,m} V_{nm} r^n Y^m_n(\theta,\phi),
\end{equation}
and equating coefficients as before
\begin{equation}
  V_{nm} = \frac{1}{a^n} S^a_{nm} = \frac{1}{b^n} S^a_{nm}
\end{equation}
so that we have for the single-layer
\begin{equation}
  \lambda^{V^L_{b,a}}_n = \left(\frac{a}{b}\right)^n \frac{a}{2n + 1}.
\end{equation}
The eigenvalues presented for these operators can be verified
analytically using Green's theorem.

\thanks{
The work of JPB was supported in part by a New Investigator award from
Rush University.
MGK acknowledges partial support from the Office of Advanced Scientific Computing Research, Office of Science, U.S. Department of Energy, under Contract DE-AC02-06CH11357
and the U.S. Army Research Laboratory and the U.S. Army Research Office under contract/grant number W911NF-09-0488.
PRB acknowledges full support from U.S. DOE Contract DE-AC01-06CH11357.
The authors gratefully acknowledge valuable discussions with A. Hildebrandt, and thank
R. S. Eisenberg for his ongoing support of their collaboration.
}

\bibliographystyle{plain}
\bibliography{implicit-review}

\begin{thebibliography}{10}

\bibitem{analytical-nonlocal-SI}
See supplementary material at [url will be inserted by aip] for matlab
  implementation of the analytical nonlocal model.

\bibitem{Ahner86}
J.~F. Ahner and R.~F. Arenstorf.
\newblock On the eigenvalues of the electrostatic integral operator.
\newblock {\em Journal of Mathematical Analysis and Applications},
  117:187--197, 1986.

\bibitem{Ahner96}
J.~F. Ahner, V.~V. Dyakin, {V. Ya. Raevskii}, and {St. Ritter}.
\newblock Spectral properties of operators of the theory of harmonic potential.
\newblock {\em Mathematical Notes}, 59(1):3--11, 1996.

\bibitem{Altman06}
M.~D. Altman, J.~P. Bardhan, B.~Tidor, and J.~K. White.
\newblock {FFTSVD}: A fast multiscale boundary-element method solver suitable
  for {BioMEMS} and biomolecule simulation.
\newblock {\em IEEE T. Comput.-Aid. D.}, 25:274--284, 2006.

\bibitem{Altman09}
M.~D. Altman, J.~P. Bardhan, J.~K. White, and B.~Tidor.
\newblock Accurate solution of multi-region continuum electrostatic problems
  using the linearized {Poisson}--{Boltzmann} equation and curved boundary
  elements.
\newblock {\em J. Comput. Chem.}, 30:132--153, 2009.

\bibitem{Atkinson97}
K.~E. Atkinson.
\newblock {\em The Numerical Solution of Integral Equations of the Second
  Kind}.
\newblock Cambridge University Press, 1997.

\bibitem{Attard90}
P.~Attard, D.~Wei, and G.~N. Patey.
\newblock Critical comments on the nonlocal dielectric function employed in
  recent theories of the hydration force.
\newblock {\em Chemical Physics Letters}, 172:69--72, 1990.

\bibitem{Azuara08}
C.~Azuara, H.~Orland, M.~Bon, P.~Koehl, and M.~Delarue.
\newblock Incorporating dipolar solvents with variable density in
  poisson--boltzmann electrostatics.
\newblock {\em Biophys. J.}, 95:5587--5605, 2008.

\bibitem{Baker01}
N.~A. Baker, D.~Sept, M.~J. Holst, and J.~A. Mc{C}ammon.
\newblock Electrostatics of nanoysystems: Application to microtubules and the
  ribosome.
\newblock {\em Proc. Natl. Acad. Sci. USA}, 98:10037--10041, 2001.

\bibitem{Bardhan11_pka}
J.~P. Bardhan.
\newblock Nonlocal continuum electrostatic theory predicts surprisingly small
  energetic penalties for charge burial in proteins.
\newblock {\em J. Chem. Phys.}, 135:104113, 2011.

\bibitem{Bardhan11_DAC}
J.~P. Bardhan and A.~Hildebrandt.
\newblock A fast solver for nonlocal electrostatic theory in biomolecular
  science and engineering.
\newblock In {\em IEEE/ACM Design Automation Conference (DAC)}, 2011.

\bibitem{Bardhan11_Knepley}
J.~P. Bardhan and M.~G. Knepley.
\newblock Mathematical analysis of the boundary-integral based electrostatics
  estimation approximation for molecular solvation: Exact results for spherical
  inclusions.
\newblock {\em J. Chem. Phys.}, 135:124107, 2011.

\bibitem{Bardhan12_Knepley_ellipsoidal}
J.~P. Bardhan and M.~G. Knepley.
\newblock Computational science and re-discovery: open-source implementation of
  ellipsoidal harmonics for problems in potential theory.
\newblock {\em Computational Science and Discovery}, 5:014006, 2012.

\bibitem{analytical-nonlocal-bitbucket}
J.~P. Bardhan, M.~G. Knepley, and P.~Brune.
\newblock Public mercurial repository containing all source code in
  supplementary material.
\newblock https://bitbucket.org/jbardhan/matlab-analytical-nonlocal-sphere.

\bibitem{Basilevsky96}
M~V Basilevsky and D~F Parsons.
\newblock An advanced continuum medium model for treating solvation effects:
  Nonlocal electrostatics with a cavity.
\newblock {\em J. Chem. Phys.}, 105(9):3734, Aug 1996.

\bibitem{Basilevsky98}
M.~V. Basilevsky and D.~F. Parsons.
\newblock Nonlocal continuum solvation model with exponential susceptibility
  kernels.
\newblock {\em J. Chem. Phys.}, 108:9107--9113, 1998.

\bibitem{Boschitsch02}
A.~H. Boschitsch, M.~O. Fenley, and H.-X. Zhou.
\newblock Fast boundary element method for the linear {P}oisson--{B}oltzmann
  equation.
\newblock {\em J. Phys. Chem. B}, 106(10):2741--54, 2002.

\bibitem{Bowman}
J.~J. Bowman, T.~B.~A. Senior, and P.~L.~E. Uslenghi.
\newblock {\em Electromagnetic and acoustic scattering by simple shapes}.
\newblock North-Holland, Amsterdam, 1969.

\bibitem{Cai11}
L.~W. Cai.
\newblock On the computation of spherical {Bessel} functions of complex
  arguments.
\newblock {\em Comp. Phys. Comm.}, 182:663--668, 2011.

\bibitem{Demchuk96}
E.~Demchuk and R.~C. Wade.
\newblock Improving the continuum dielectric approach to calculating {pKas} of
  ionizable groups in proteins.
\newblock {\em J. Phys. Chem.}, 100:17373--17387, 1996.

\bibitem{Dogonadze74}
R.~R. Dogonadze and A.~A. Kornyshev.
\newblock Polar solvent structure in the theory of ionic solvation.
\newblock {\em J. Chem. Soc. Faraday Trans. 2}, 70:1121--1132, 1974.

\bibitem{Engelen03}
R.~A.~B. Engelen, M.~G.~D. Geers, and F.~P.~T. Baaijens.
\newblock Nonlocal implicit gradient-enhanced elasto-plasticity for the
  modelling of softening behavior.
\newblock {\em International Journal of Plasticity}, 19:403--433, 2003.

\bibitem{Fasel08}
C.~Fasel, S.~Rjasanow, and O.~Steinbach.
\newblock A boundary integral formulation for nonlocal electrostatics.
\newblock In Karl Kunisch, Günther Of, and Olaf Steinbach, editors, {\em
  Numerical Mathematics and Advanced Applications}, pages 117--124. Springer
  Berlin Heidelberg, 2008.

\bibitem{Gilson86}
M.~Gilson and B.~Honig.
\newblock The dielectric constant of a folded protein.
\newblock {\em Biopolymers}, 25:2097--2119, 1986.

\bibitem{Gong08}
H.~Gong, G.~Hocky, and K.~F. Freed.
\newblock Influence of nonlinear electrostatics on transfer energies between
  liquid phases: Charge burial is far less expensive than {Born} model.
\newblock {\em Proc. Natl. Acad. Sci. USA}, 105:11146--11151, 2008.

\bibitem{Havranek99}
J.~J. Havranek and P.~B. Harbury.
\newblock {Tanford}--{Kirkwood} electrostatics for protein modeling.
\newblock {\em Proc. Natl. Acad. Sci. USA}, 96(20):11145--11150, 1999.

\bibitem{Helland99}
R.~Helland, J.Otlewski, O.~Sundheim, M.~Dadlez, and A.~O.Smalas.
\newblock The crystal structures of the complexes between bovine beta-trypsin
  and ten {P1} variants of {BPTI}.
\newblock {\em J. Mol. Biol.}, 287:923--942, 1999.

\bibitem{Hildebrandt05}
A.~Hildebrandt.
\newblock {\em Biomolecules in a structured solvent: A novel formulation of
  nonlocal electrostatics and its numerical solution}.
\newblock PhD thesis, Universit\"{a}t des Saarlandes, 2005.

\bibitem{Hildebrandt04}
A.~Hildebrandt, R.~Blossey, S.~Rjasanow, O.~Kohlbacher, and H.-P. Lenhof.
\newblock Novel formulation of nonlocal electrostatics.
\newblock {\em Phys. Rev. Lett.}, 93:108104, 2004.

\bibitem{Hildebrandt07}
A~Hildebrandt, R~Blossey, S~Rjasanow, O~Kohlbacher, and H.-P Lenhof.
\newblock Electrostatic potentials of proteins in water: a structured continuum
  approach.
\newblock {\em Bioinformatics}, 23(2):e99--e103, Jan 2007.

\bibitem{Hsiao94}
G.~C. Hsiao and R.~E. Kleinman.
\newblock Error analysis in numerical solution of acoustic integral equations.
\newblock {\em International Journal for Numerical Methods in Engineering},
  37:2921--2933, 1994.

\bibitem{Jackson_classical_electrodynamics}
J.~D. Jackson.
\newblock {\em Classical Electrodynamics}.
\newblock Wiley, 3$^{rd}$ edition, 1998.

\bibitem{Juffer91}
A.~H. Juffer, E.~F.~F. Botta, B.~A.~M. van Keulen, A.~van~der Ploeg, and
  H.~J.~C. Berendsen.
\newblock The electric potential of a macromolecule in a solvent: A fundamental
  approach.
\newblock {\em J. Comput. Phys.}, 97(1):144--171, 1991.

\bibitem{Kangas98}
E.~Kangas and B.~Tidor.
\newblock Optimizing electrostatic affinity in ligand--receptor binding:
  Theory, computation, and ligand properties.
\newblock {\em J. Chem. Phys.}, 109:7522--7545, 1998.

\bibitem{Kirkwood34}
J.~G. Kirkwood.
\newblock Theory of solutions of molecules containing widely separated charges
  with special application to zwitterions.
\newblock {\em J. Chem. Phys.}, 2:351, 1934.

\bibitem{Ritter95_spectrum}
R.~E. Kleinman, R.~Kress, and E.~Martensen, editors.
\newblock {\em The spectrum of the electrostatic integral operator for an
  ellipsoid}, Frankfurt/Bern, 1995. Lang.

\bibitem{Koehl10}
P.~Koehl and M.~Delarue.
\newblock {AQUASOL}: an efficient solver for the dipoar
  {Poisson}--{Boltzmann}--{Langevin} equation.
\newblock {\em J. Chem. Phys.}, 132:064101, 2010.

\bibitem{Kornyshev78}
A.~A. Kornyshev, A.~I. Rubinshtein, and M.~A. Vorotyntsev.
\newblock Model nonlocal electrostatics: I.
\newblock {\em Journal of Physics C: Solid State Physics}, 11:3307, Dec 1978.

\bibitem{Kuo02}
S.~S. Kuo, M.~D. Altman, J.~P. Bardhan, B.~Tidor, and J.~K. White.
\newblock Fast methods for simulation of biomolecule electrostatics.
\newblock In {\em International Conference on Computer Aided Design (ICCAD)},
  2002.

\bibitem{Liang97}
J.~Liang and S.~Subramaniam.
\newblock Computation of molecular electrostatics with boundary element
  methods.
\newblock {\em Biophys. J.}, 73(4):1830--1841, 1997.

\bibitem{Lu06}
B.~Z. Lu, X.~L. Cheng, J.~Huang, and J.~A. {McCammon}.
\newblock Order {N} algorithm for computation of electrostatic interactions in
  biomolecular systems.
\newblock {\em Proc. Natl. Acad. Sci. USA}, 103(51):19314--19319, 2006.

\bibitem{Mladenov96}
I.~M. Mladenov.
\newblock {Kirkwood's} formula revisited.
\newblock {\em Europhysics Letters}, 33:577--581, 1996.

\bibitem{Ochs98}
R.~L. {Ochs Jr.} and G.~Kristensson.
\newblock Using local differential operators to model dispersion in dielectric
  media.
\newblock {\em Journal of the optical society of America A}, 15:2208--2215,
  1998.

\bibitem{Roux99}
B.~Roux and T.~Simonson.
\newblock Implicit solvent models.
\newblock {\em Biophys. Chem.}, 78:1--20, 1999.

\bibitem{Rubinstein10}
A~Rubinstein, R~Sabirianov, W~Mei, F~Namavar, and A~Khoynezhad.
\newblock Effect of the ordered interfacial water layer in protein complex
  formation: A nonlocal electrostatic approach.
\newblock {\em Phys. Rev. E}, 82(2):021915, Aug 2010.

\bibitem{Rubinstein04}
Alexander Rubinstein and Simon Sherman.
\newblock Influence of the solvent structure on the electrostatic interactions
  in proteins.
\newblock {\em Biophys. J.}, 87(3):1544--1557, Sep 2004.

\bibitem{Rubinstein07}
Alexander Rubinstein and Simon Sherman.
\newblock Evaluation of the influence of the internal aqueous solvent structure
  on electrostatic interactions at the protein-solvent interface by nonlocal
  continuum electrostatic approach.
\newblock {\em Biopolymers}, 87(2-3):149--164, Oct 2007.

\bibitem{Schutz01}
C.~N. Schutz and A.~Warshel.
\newblock What are the dielectric constants of proteins and how to validate
  electrostatic models?
\newblock {\em Proteins}, 44:400--417, 2001.

\bibitem{Sham98}
Y.~Y. Sham, I.~Muegge, and A.~Warshel.
\newblock The effect of protein relaxation on charge-charge interactions and
  dielectric constants of proteins.
\newblock {\em Biophys. J.}, 74(4):1744--1753, 1998.

\bibitem{Sigalov06}
G.~Sigalov, A.~Fenley, and A.~Onufriev.
\newblock Analytical electrostatics for biomolecules: {Beyond} the generalized
  {Born} approximation.
\newblock {\em J. Chem. Phys.}, 124(124902), 2006.

\bibitem{Sigalov05}
G.~Sigalov, P.~Scheffel, and A.~Onufriev.
\newblock Incorporating variable dielectric environments into the generalized
  {Born} model.
\newblock {\em J. Chem. Phys.}, 122:094511, 2005.

\bibitem{Sitkoff94}
D.~Sitkoff, K.~A. Sharp, and B.~Honig.
\newblock Accurate calculation of hydration free energies using macroscopic
  solvent models.
\newblock {\em J. Phys. Chem. B}, 98:1978--1988, 1994.

\bibitem{VanBladel}
J.~{Van Bladel}.
\newblock {\em Electromagnetic Fields}.
\newblock John Wiley \& Sons, Hoboken, NJ, second edition, 2007.

\bibitem{Vorotyntsev78}
M.~A. Vorotyntsev.
\newblock Model nonlocal electrostatics: Ii. spherical interface.
\newblock {\em J. Phys. C: Solid State Phys.}, 11:3323--3331, 1978.

\bibitem{Warwicker82}
J.~Warwicker and H.~C. Watson.
\newblock Calculation of the electric potential in the active site cleft due to
  alpha-helix dipoles.
\newblock {\em J. Mol. Biol.}, 157:671--679, 1982.

\bibitem{Weggler_thesis}
S.~Weggler.
\newblock {\em Correlation induced electrostatic effects in biomolecular
  systems}.
\newblock PhD thesis, Universit\"{a}t des Saarlandes, 2010.

\bibitem{Weggler10}
S~Weggler, V~Rutka, and A~Hildebrandt.
\newblock A new numerical method for nonlocal electrostatics in biomolecular
  simulations.
\newblock {\em J. Comput. Phys.}, 229(11):4059--4074, Jun 2010.

\bibitem{Yoon90}
B.~J. Yoon and A.~M. Lenhoff.
\newblock A boundary element method for molecular electrostatics with
  electrolyte effects.
\newblock {\em J. Comput. Chem.}, 11(9):1080--1086, 1990.

\bibitem{Zhou97_2}
H.-X. Zhou.
\newblock Control of reduction potential by protein matrix: lesson from a
  spherical protein model.
\newblock {\em Journal of Biological Inorganic Chemistry}, 2:109--113, 1997.

\end{thebibliography}

\end{document}